# Fabrication of polarization-independent waveguides deeply buried in crystal using aberration-corrected femtosecond laser direct writing

Peng Wang[1,2,3], Jia Qi[1,2,3], Zhengming Liu[1,2,3], Yang Liao[1], Wei Chu[1], Ya Cheng[1,4,5]

[1] State Key Laboratory of High Field Laser Physics, Shanghai Institute of Optics and Fine Mechanics, Chinese Academy of Sciences, Shanghai 201800, China.
[2] School of Physical Science and Technology, ShanghaiTech University, Shanghai 200031, China.
[3] University of Chinese Academy of Sciences, Beijing 100049, China.
[4] State Key Laboratory of Precision Spectroscopy, East China Normal University, Shanghai 200062, China.
[5] Collaborative Innovation Center of Extreme Optics, Shanxi University, Taiyuan, Shanxi 030006, China.

Correspondence and requests for materials should be addressed to Y. L. (email: superliao@vip.sina.com) or Y.C. (email: ya.cheng@siom.ac.cn).

**ABSTRACT**

Writing optical waveguides with femtosecond laser pulses provides the capability of forming three-dimensional photonic circuits for manipulating light fields in both linear and nonlinear manners. To fully explore this potential, large depths of the buried waveguides in transparent substrates are often desirable to facilitate achieving vertical integration of waveguides in a multi-layer configuration, which, however, is hampered by rapidly degraded axial resolution caused by optical aberration. Here, we show that with the correction of the spherical aberration, polarization-independent waveguides can be inscribed in a nonlinear optical crystal lithium niobate (LN) at depths up to 1400 μm, which is more than one order of magnitude deeper than the waveguides written with aberration uncorrected femtosecond laser pulses. Our technique is beneficial for applications ranging from miniaturized nonlinear light sources to quantum information processing.

**Introduction**

Integrated photonic devices have found broad range of applications including information processing, chemical and biological analyses thanks to their diverse functionalities and high operation efficiencies [1, 2]. One of the key elements of photonic integration is the optical waveguide, which provides predefined pathways for transferring light signals among different nodes in optical circuits. Furthermore, since the optical waveguides can confine the light fields in a tight space typically on the scale of wavelength of light, interaction of light with the optical materials can be enhanced in the waveguide, giving rise to enhanced efficiency of nonlinear frequency conversion [3]. These benefits make it attractive to build quantum photonic chips based on producing waveguides in nonlinear crystals such as LN and diamond [4, 5].

Recently, it has been demonstrated that formation of optical waveguides can be efficiently achieved by femtosecond laser direct writing in transparent materials [6-10]. Writing waveguides with focused femtosecond laser pulses offers a continuous single-step processing approach with the capability of producing arbitrary three-dimensional (3D) optical circuits, which is desirable for miniaturization and integration [11-13]. Specifically, inscription of waveguides in nonlinear crystals has attracted great attention as this technique holds the promising potential for on-chip nonlinear optical applications and quantum information processing. Toward this end, both double-line waveguides [9] and depressed cladding waveguides [14] have been produced in crystals using femtosecond laser direct writing. The former type waveguides are easy to fabricate, whereas such waveguides show strong polarization dependence owing to the fact that the top and bottom claddings of the waveguides are missing [15]. The latter type of waveguides can guide the light waves polarized in both horizontal and vertical directions, whereas forming such depressed cladding waveguides can often be a time consuming process which requires overlapping large numbers of lines inscribed by the focused femtosecond laser to form the enclosed cladding [16].

To overcome the above difficulties, we have demonstrated fabrication of polarization-independent waveguides in crystal by forming square-shaped depressed cladding (SSDC) [17]. Unlike the conventional depressed cladding waveguides with a circular contour, our SSDC waveguides are composed of only four sides. Therefore, only four scans are required to form the SSDC waveguides regardless of the mode-field size. In addition, since the SSDC waveguides can be inscribed with low-numerical aperture lenses, the waveguides can be buried deeply in the substrates to facilitate producing large-scale multi-layered 3D photonic circuits. In this work, we show that by correcting the aberration in the focal system, we are able to produce SSDC waveguides in LN crystal which supports both s- and p- polarization modes at a depth of 1400 μm. The measured insertion losses are comparable for both s- and p- polarization modes.

**Results and Discussion**

In our experiment, commercially available MgO-doped x-cut $LiNbO_3$ crystals of a size of 10×5×3 $mm^3$ were used as the substrates. Figure 1 shows the schematic of our experimental setup by which adaptive slit beam shaping was achieved by a spatial light modulator (SLM) [18]. The key to produce the SSDC waveguides is to write homogeneous horizontal lines with heights (i.e., the sizes of the lines in the vertical direction) of a couple of microns and lengths (i.e., the sizes of the lines in the horizontal direction) ranging from a few microns to nearly twenty microns, depending on the mode-field size demanded by the applications. Writing of the vertical lines can be carried out in a straightforward manner, as the heights of these lines can be tuned by choosing the NA of focal lens and the pulse energy. The NA of focal lens determines the linear focal depth; whereas the pulse energy will further extend the focal depth beyond that allowed by the linear focusing due to the nonlinear self-focusing. For this reason, we compare two horizontal lines written at a depth of 40 μm in the LN substrate with and without use of the aberration correction, as shown in Figs. 2(a) and (b), respectively. In writing of the lines in Figs. 2(a) and (b), the average power was set at 2 mW, and the scan speed was set at 20 μm/s. The phase mask for writing the horizontal and vertical lines are illustrated in Fig. 1. The correction ring of objective lens (Olympus LUCPLFLN,

NA = 0.7,) was tuned to match the depth of 40 μm in LN crystal, which has an ordinary refractive index of 2.25 and an extraordinary refractive index of 2.17. It can be seen that at such a depth, the aberration is already too severe to allow for writing the horizontal lines in the LN crystal without the aberration correction. The peak intensity is already high enough to damage the surface of the LN substrate, as shown in Fig. 2(a). In addition, the internal structure produced below the surface of LN substrate has an irregular shape elongated along the axial direction resulting from the interplay between the aberration and self-focusing. However, with the aberration correction, a homogeneous horizontal line with a height of 3 μm and length of 16 μm was produced, as shown in Fig. 2(b). Such horizontal lines can be used for constructing polarization-independent waveguides in the LN crystal as reported in our previous work [17].

We then carried out the same experiment but only changed the depth of the buried horizontal lines from 40 μm to 120 μm. The correction ring of objective lens was tuned accordingly. We set the average power to 3 mW and the scan speed maintained to be 20 μm/s. As shown in Fig. 3(a), at this moderate depth, only weak modification occurred in the LN crystal owing to the spherical aberration. The laser modified region consists of several discrete lines along the propagation direction, indicating the occurrence of multiple filamentation. In contrast, the optical micrograph in Fig. 3(b) clearly shows that after the correction of the aberration, a nice horizontal line with a depth of 3 μm and a length of 16 μm was produced at the depth of 120 μm. The cross-section of a SSDC waveguide constructed with the line in Fig. 3(b) is shown in Fig. 3(c), and the near-field profiles of the guided modes of s- and p- polarization are displayed in Figs. 3(d) and 3(e), respectively. To produce the waveguide in Fig. 3(c), the paired vertical lines were written with the same objective lens whereas a different phase mask was chosen for the beam shaping [see methods]. The vertical lines were written in the LN crystal at an average power of 1 mW and scan speed of 50 μm/s. The insertion losses of the waveguide in Fig. 3(c) were measured to be 4.5 dB and 4 dB for p- and s- polarization. It should be mentioned that the insertion loss includes a coupling loss caused by mode-field mismatching between the waveguide and the optical fiber and a loss caused by the Fresnel reflections at the two facets of the waveguide. Therefore, the propagation loss in the waveguide should be much less than the insertion losses measured above. The necessity of using aberration correction in writing the SSDC waveguides in crystals is unambiguously proved.

In our experiment, we confirm that after using the aberration correction, homogeneous horizontal lines can even be written at a depth of 1400 μm, as shown in Fig. 4(a). The produced horizontal line has a height of 3 μm and a length of 18 μm. An average power of 6 mW and a scan speed of 20 μm/s were chosen for writing the horizontal lines. Then, we constructed a SSDC waveguide using the horizontal line in Fig. 4(a), which is displayed in Fig. 4(b). Here, the vertical lines were written in the LN crystal at an average power of 1.3 mW and scan speed of 50 μm/s. The near-field mode profiles of the s- and p-polarized light are shown in Figs. 4(c) and 4(d), respectively. The insertion losses of the waveguide were measured to be 3.9 dB and 3.2 dB for the p- and s-polarized light, respectively. The successful demonstration of the polarization-independent waveguides at such a great depth will facilitate high-density integration of complex multi-layered circuits to increase the capacity of photonic microdevices.

## Conclusion

To conclude, we have demonstrated that depressed cladding waveguides, which can allow polarization-independent optical waveguiding in crystals, can be written deeply inside LN with a depth of 1400 μm. This is enabled by use of shaped femtosecond laser pulses to produce SSDC waveguides with a low-NA, long-working distance focal lens. We show that aberration correction is of vital importance to achieve this goal as many crystals have higher refractive indices than that of glass materials, thus the spherical aberration will induce severe elongation of the focal spot in the axial direction. The degraded longitudinal resolution is further affected by nonlinear self-focusing at high laser intensities such as that used for femtosecond laser processing. Our result provides clear evidence that the SSDC waveguides will have much use for constructing 3D nonlinear optical or quantum photonic circuits in various crystalline substrates.

## Methods

In our experiment, commercially available MgO-doped x-cut LiNbO$_3$ crystals of a size of 10× 5 × 3 mm$^3$ were used as the substrates. Figure 1 shows the schematic of our experimental setup. The output beam of a Ti:Sapphire laser (Libra-HE, Coherent Inc.) with a maximum pulse energy of 3.5 mJ, an operation wavelength of 800 nm, a pulse width of ~50 fs, and a repetition rate of 1 kHz was used. The pulse energy was controlled by a rotatable half-wave plate and a Glan-Taylor polarizer. The laser beam was first expanded using a concave lens of focal length 7.5 cm (L1) and a convex lens of focal length 15 cm (L2) before impinging on a reflective phase-only spatial light modulator (SLM, Hamamatsu, X10468-02). To form the slit-shaped beam with the phase-only SLM, a blazed grating with a modulation depth of 2π rad and a period of 420 μm was used to maximize the diffraction efficiency of the first order. Example phase masks used for writing the horizontal and vertical sides of the cladding are presented in the insets, respectively. The horizontal lines were produced by setting the length and width of the slit on the SLM to 12 mm and 0.28 mm, respectively; whereas the vertical lines were written using a slit with a length of 5.6 mm and a width of 5.2 mm, respectively. The first diffracted order produced by the SLM, after being focused with a lens of focal length 150 cm (L3), was filtered out using a pinhole. The slit-shaped beam were then imaged onto the back aperture of the objective lens (Olympus LUCPLFLN 60×, NA = 0.7, WD = 1.5 ~ 2.2 mm) using a lens of focal length 50 cm (L4), and focused into the glass sample to write the horizontal and vertical claddings of buried waveguides. When writing at the different depths, aberration correction matched to the glass thickness is accomplished using a correction ring. The sample was translated along the crystal y-direction by a computer-controlled XYZ stage with a translation resolution of 1 μm. For waveguide characterization, a fiber laser with a wavelength of 1550 nm (THORLABS S1FC1550) was used for end-fire coupling into the waveguide. To investigate the polarization dependence in the fabricated waveguides, a polarization maintaining fiber (PMF) is used to control the polarization direction of the input beams. The outcoming beam is collimated using a 20× objective of NA = 0.45. The near–field mode profile at the exit facet is imaged onto an InGaAs camera (HAMAMATSU C12741-03) with the objective lens. The power of the collimated beam is measured using a calibrated mid-infrared detector. Insertion losses of all the presented waveguides are obtained by dividing the measured output powers from the SSDC waveguides by the input laser powers.

**Figure Captions**

Figure 1. Schematic illustration of the experimental setup. HWP: half-wave plate. POL: polarizer. P: pinhole. CCD: charge coupled device. OBJ: objective lens. PC: personal computer. L1, L2, L3, and L4 are the lenses of different focal lengths which are described in the main text. Coordinates are indicated in the figure. Inset: example phase masks for writing the horizontal and vertical sides of the cladding, respectively.

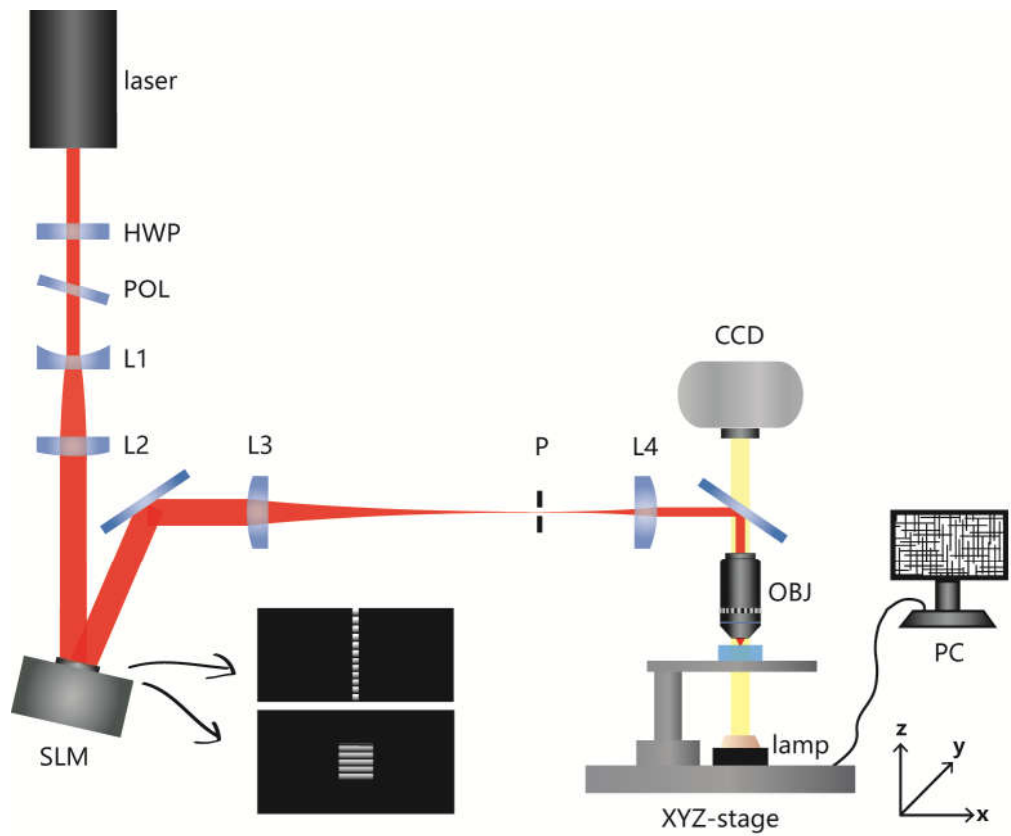

Figure 2. Optical micrograph of laser-affected zone at a depth of 40 μm. (a) Without aberration correction. (b) With aberration correction.

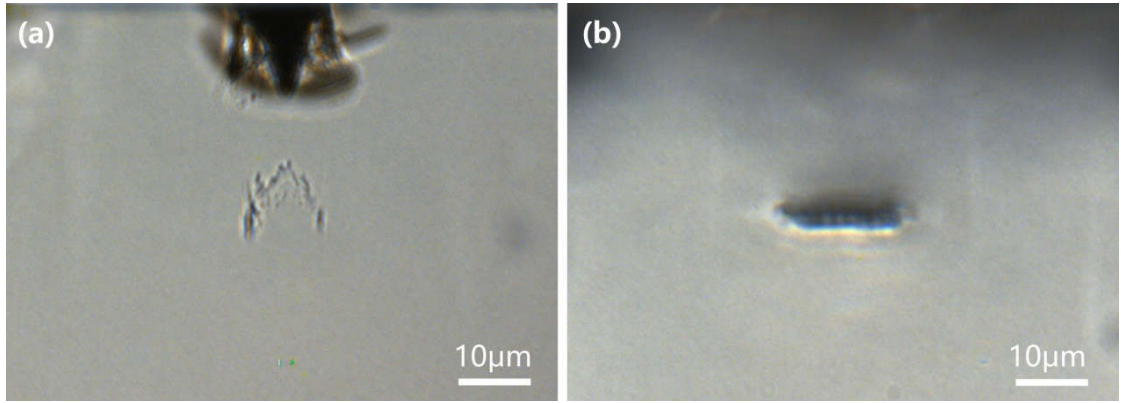

Figure 3. Optical micrograph of laser-affected zone at a depth of 120 μm. (a) Without aberration correction. (b) With aberration correction. (c) Optical micrograph of the cross section of a SSDC waveguide fabricated in LN crystal at the same depth. (d) Near-field mode profile of s-polarized beam in the waveguide. (e) Near-field mode profile of p-polarized beam in the waveguide.

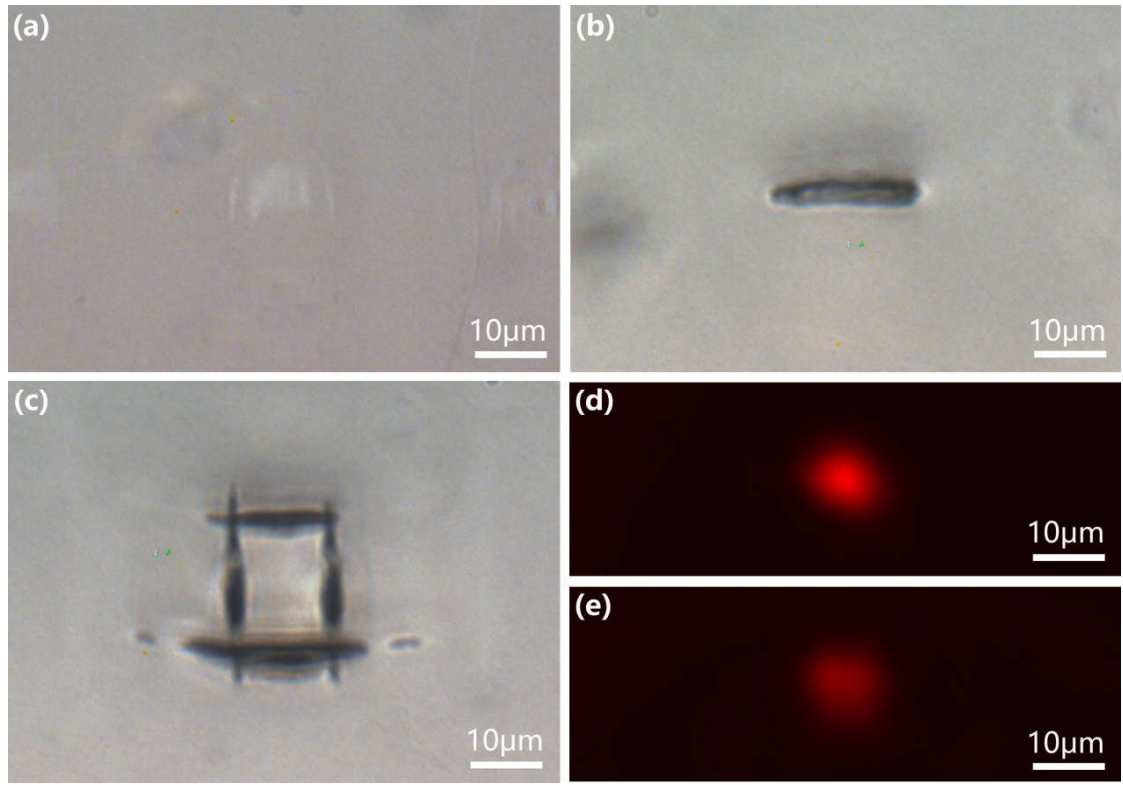

Figure 4. (a) Optical micrograph of horizontal line written at a depth of 1400 μm with aberration correction. (b) Optical micrograph of the cross section of a SSDC waveguide fabricated in LN crystal at the same depth. (d) Near-field mode profile of s-polarized beam in the waveguide. (e) Near-field mode profile of p-polarized beam in the waveguide.

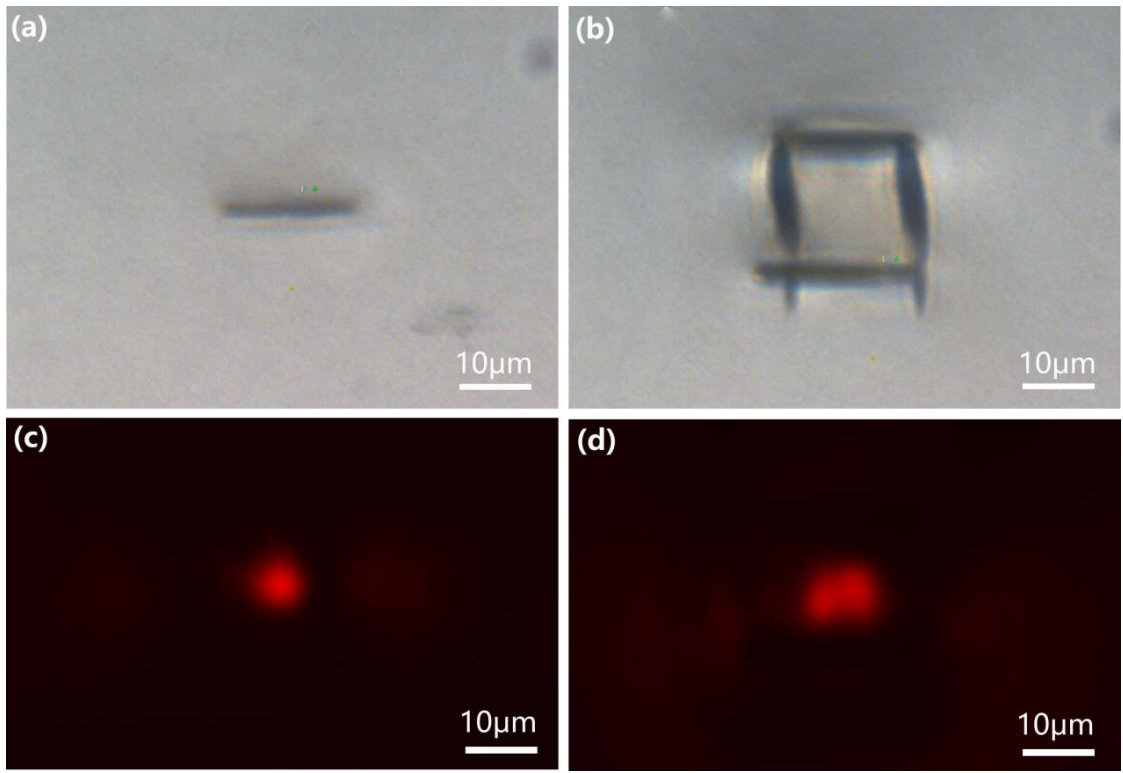